# Strategies to simulate dephasing-assisted quantum transport on digital quantum computers

Federico Gallina[1], Matteo Bruschi[1], Barbara Fresch[1,2]

[1] Dipartimento di Scienze Chimiche, Università degli Studi di Padova, via Marzolo 1, Padua 35131, Italy

[2] Padua Quantum Technologies Research Center, Università degli Studi di Padova

## Abstract

Simulating charge and energy transfer in extended molecular networks requires an effective model to include the environment because it significantly affects the quantum dynamics. A prototypical effect known as Environment-Assisted Quantum Transport (ENAQT) consists in the enhancement of the transfer efficiency by the interaction with an environment. A simple description of this phenomenon is obtained by a quantum master equation describing a quantum walk over the molecular network in the presence of inter-site decoherence. We consider the problem of simulating the dynamics underlying ENAQT in a digital quantum computer. Two different quantum algorithms are introduced, the first one based on stochastic Hamiltonians and the second one based on a collision scheme. We test both algorithms by simulating ENAQT in a small molecular network on a quantum computer emulator and provide a comparative analysis of the two approaches. Both algorithms can be implemented in a memory efficient encoding with the number of required qubits scaling logarithmically with the size of the simulated system while the number of gates increases quadratically. We discuss the algorithmic quantum trajectories generated by the two simulation strategies showing that they realize distinct *unravellings* of the site-dephasing master equation. In our approach, the non-unitary dynamics of the open system is obtained through effective representations of the environment, paving the way to digital quantum simulations of quantum transport influenced by structured environments.

# 1 Introduction

A large variety of mathematical and physical problems can be translated into a graph and solved by a walker moving randomly from a node to another connected node. Energy and charge transfer processes in molecular networks are prominent examples. Random walks define a family of probabilistic models providing a general paradigm for sampling and exploring extended networks by using a sequence of simple local transitions. Notably, the Continuous-Time Quantum Walk (QW) was early introduced as the quantum analogue of the classical random walk algorithm [1] and it assures a quantum advantage in several computational tasks like traversing decision trees [1,2], searching unstructured databases [3,4] and solving hard satisfiability problems [5]. The continuous-time quantum walk is defined by building a quantum Hamiltonian according to the topology of the underlying graph, where the walker moves among the nodes following the rules of quantum dynamics. Important results by Childs *et al.* highlighted that quantum walk itself is a universal computational primitive [6], meaning that any problem that can be handled by a general-purpose quantum computer can be translated into a quantum walk over a graph. By its definition, QW computation is intimately related to the dynamics of quantum transport and it offers an efficient simulation routine for the coherent transport over a network of quantum states [7]. The relation also applies in the opposite direction, and QW algorithms can be implemented by the dynamics of isolated and controllable physical systems which act as analog simulators. Quantum walkers have been experimentally realized with trapped ions [8], phonons of trapped ions [9], neutral atoms in optical traps [10] and excitation in superconducting circuits [11]. In the digital setting of gate-based quantum computers, the implementation of QW is still challenging as it requires circuits of a significant depth. However, the algorithms developed to simulate efficiently quantum dynamics can be used to implement QW routines as the quality of digital quantum hardware improves.

Because the exponential speedup in traversing certain graph topologies arises from constructive interference effects, one might expect that the purely unitary quantum transport described by the QW algorithm is the fastest and the most efficient way to propagate through a network of quantum sites. However, such an expectation turns out to be wrong. Indeed, a rather general scenario is that the presence of a dephasing environment coupled to the system can enhance quantum transport, in a phenomenon named Environment-Assisted Quantum Transport [12] (ENAQT). Although ENAQT can occur on ordered lattices [12,13], transport enhancement was first characterized in disordered systems in the context of exciton transfer in light-harvesting complexes. In the first step of the photosynthesis, after the antenna chromophores absorbed a photon, an exciton is formed that is usually strongly coupled with intramolecular vibrations while spreading over the molecular complex to reach the reaction centre. The high efficiency of the energy transfer and the long-living coherent beatings that were experimentally measured [14–16] triggered a lively debate on the interplay between Hamiltonian dynamics, static disorder and decoherence in affecting the efficiency of the energy transport. Substantial theoretical work demonstrated that an intermediate level of environmental-induced dephasing can indeed explain the high efficiency and partial coherent character of the transport [17,18]. Mohseni *et al.* formalized the environment-assisted quantum transport in the FMO antenna complex in terms of a generalized model for quantum walks which includes dephasing and dissipation effects into the dynamics of the walker [19]. The basic phenomenology of ENAQT is described by a Lindblad master equation in which the Hamiltonian describes the coupling between different quantum sites and local Lindblad terms accounts for the dephasing induced by the surrounding environment. Recent quantum simulations of ENAQT on analog quantum devices (photonic setups [20,21] ultracold atoms [22], superconducting circuits [23–25], nuclear spin systems [26] and trapped atoms [27]) confirmed the theoretical predictions and offer a controlled environment to further test the effects of noise and interactions in quantum transport.

In this work, we consider the problem of simulating dephasing ENAQT on a gate-based (digital) quantum computer and we explicitly formulate, discuss and test two different algorithms. Our motivation and the relevance of the results are twofold: firstly, simulation of the transport efficiency in complex networks have an important impact on technologies ranging from synthetic light-harvesting systems to chemical sensors, to photovoltaic, where the aim is to design materials and engineering environments to maximize efficiency [21]. Taking advantage of the favourable scaling of memory requirements, quantum simulations of mesoscopic chromophore networks might be possible on quantum computers. On the other hand, ENAQT represents a moderately sophisticated quantum dynamical process, and the comparison of different quantum simulation algorithms offers a fresh view of the underlying mechanism. In particular, we will discuss how different algorithms provide different "unravellings" of the average evolution corresponding to the solution of the site-dephasing master equation.

The process of site-decoherence is physically motivated by the interaction of the quantum sites with an external dephasing environment. To study the interplay of unitary quantum evolution and decoherence in digital quantum simulations we need to face the inherent difficulty of mapping non-unitary evolution into the framework of unitary gates. General strategies to tackle this problem have been developed, mostly based on dilation of the Hilbert space of the system [28,29] and variational principles [30]. Dilation of the Hilbert space allows one to obtain the non-unitary dynamics of the system of interest by simulating the unitary dynamics of a larger system. The main problem of this approach is that the global unitary dynamics is not necessarily ruled by a physical (local) Hamiltonian and therefore it may not be simulated efficiently. Variational techniques are promising for implementation in near-term devices as they usually require shallow circuits, however, they are heavily affected by the quality of the classical optimization procedure [31]. Within this framework, few digital quantum algorithms for the simulation of environment-assisted quantum transport have been recently reported, focusing on the case study of exciton transport in molecular networks: in ref. [32], Mahdian *et al.* analysed the ENAQT master equation of the FMO antenna complex by decomposing the unitary and non-unitary part of the generator of the ENAQT dynamics into generators belonging to the universal set discussed in [33] and propose a circuital implementation on NMR quantum computer [34]. Gupta and Chandrashekar cast the solution of the ENAQT dynamics in the form of an operator sum representation for the evolution of the density matrix during discrete time steps and sketched a quantum algorithm that implements the coherent dynamics and the transitions between different excitonic states in the FMO complex [35]. Finally, in ref. [36], Hu *et al.* implemented the operator sum representation of the ENAQT master equation by the dilation approach presented in ref. [29] also discussing results obtained with the IBM Qiskit QASM simulator. Notice that these approaches require formulating the dynamics of the open system in terms of a set of *quantum channels* mapping the initial reduced density matrix to the reduced density matrix at subsequent times. Translating the differential master equation into a sequence of quantum channels is a necessary preliminary step that poses already a non-trivial challenge [37]. In general, there is not a straightforward route to identify the proper quantum channels without knowing the microscopic details of the environment.

In this paper, we design and test two quantum algorithms for gate-based simulation of dephasing ENAQT inspired by the underlying idea of devising *effective* representations of the environment. The long-term goal of this approach is to be able to "perform experiments" through the quantum computer, meaning to simulate open system dynamics which are not known *a priori* but rather generated in the computer by explicitly controlling the features of the effective environment. Following the stochastic approach to exciton dynamics pioneered by Haken and Strobl [38], in the first algorithm the environment is represented by a stochastic noise in the Hamiltonian of the system, we will refer to this scheme as the *classical noise algorithm*. The second algorithm is based on a quantum collisional model [39] and we will show how a single ancilla qubit is sufficient to encode the dephasing role of the environment, we will refer to this implementation as the *collision algorithm.* We test both algorithms by demonstrating ENAQT in a system of four sites with nearest-neighbour interactions (cyclic topology network). Our algorithms differ from previous proposals [32,34–36] both in terms of the approach and in terms of scalability which will be discussed in some detail. We expect that, relying on the somehow explicit representation of the environment, the effects of finite temperature, coloured noises and non-Markovian evolution can be easily added to our algorithms paving the way to digital quantum simulations of transport influenced by structured environments.

The paper is organized as follows: in Section 2, we introduce ENAQT in terms of a Lindblad master equation corresponding to the Haken-Strobl model for exciton diffusion. The Hamiltonian part of the model accounts for the coupling between different quantum sites while the role of the environment is to induce decoherence between different sites. We define the transport efficiency and discuss how the molecular network can be encoded in the quantum computer. Section 3 is dedicated to the development of the classical noise algorithm based on a stochastic Hamiltonian with diagonal disorder. The solution of the master equation is obtained by an explicit average of the Schrödinger dynamics corresponding to different realizations of the noise. In Section 4, we discuss an algorithm based on a collision model and we demonstrate how a single ancilla qubit is sufficient to encode the effect of the environment. Both algorithms can be run encoding each site of the network in a qubit (physical mapping) or by encoding each site of the network into an element of the computational basis (algorithmic mapping). A comparative analysis of the computational resources required by the two algorithms using the two mappings is detailed in Section 5. The analysis is supplemented with the results obtained by running the algorithms on a quantum computer emulator, namely the IBM Qiskit QASM simulator [40]. Because of the necessity of repeating the quantum circuit to accumulate measurement statistics, each execution of the circuit realizes a quantum trajectory, which we analyse in Section 6, followed by conclusions and perspectives.

## 2 Dephasing-assisted quantum transport in a quantum computer

Before tackling the issue of quantum simulations, let us frame the relation between the quantum walk on a graph and the problem of quantum transport through molecular networks where the introduction of a dephasing environment is naturally considered. The graph is a structure made of a set of vertices connected by edges according to a certain topology, which can be described by the Laplacian matrix

$$\mathcal{L} = \sum_{a=1}^{N} d_a |a\rangle\langle a| - \sum_{\langle a,b \rangle} \left( |a\rangle\langle b| + |b\rangle\langle a| \right), \quad (1)$$

where $N$ is the number of vertices of the graph represented by the basis $|a\rangle$, $d_a$ is the number of links to other vertices (the degree of vertex $a$) and $\langle a,b \rangle$ indicates an edge between two connected vertices. The basic idea of Farhi and Gutmann [1] to define a Quantum Walk on the graph is to use the Laplacian matrix in eq. (1) to specify a quantum Hamiltonian, $H = -\nu\mathcal{L}$, where coefficient $\nu$ sets the rate of transport. The dynamics of the quantum walker initially placed at node $a$ is then ruled by the corresponding Schrödinger equation, and the probability to go from node $a$ at $t=0$ to node $b$ at time $t$ is thus defined as

$$p(b,t|a) = \left| \langle b | \exp(-iHt) | a \rangle \right|^2, \quad (2)$$

where we set $\hbar = 1$. From these definitions, any physical system whose Hamiltonian can be cast in the required form realises an instance of the QW scheme. A straightforward analogy with eq. (1) can be found in the tight-binding Hamiltonian used in solid-state physics [41] and the Hückel model for molecular orbitals [42], but also with the Frenkel exciton Hamiltonian used to model excitation on chromophore networks [43] and in general with model Hamiltonian built to describe charge and energy transfer in weakly interacting molecular aggregates and nanostructures [44]. In all these cases, we define a set of quantum sites $\{j\}$. When there is one quantum state at each site, the Hamiltonian can be written as

$$H = \sum_{j=1}^{N} \varepsilon_j |j\rangle\langle j| + \sum_{\langle j,j' \rangle} V_{j,j'} \left( |j\rangle\langle j'| + |j'\rangle\langle j| \right), \quad (3)$$

with $\varepsilon_j$ being the site energy and $V_{j,j'}$ the coupling constant between two sites. By comparing eq. (3) with eq. (1) we can see that in principle we can implement a QW on, *e.g.*, a cyclic lattice by considering a ring of tunnel coupled quantum dots [45] or a chromophore ring of the LH2 antenna complex [46]. However, since nanostructures are rarely free of imperfections and molecular systems have a multitude of degrees of freedom, the effects of static and dynamic disorder into the dynamics of the quantum walk become of central interest. A common feature is the presence of (static) energy disorder in the diagonal terms of the Hamiltonian. If disorder is high, quantum transport can be suppressed by Anderson localization with detrimental effects on the quantum advantage provided by the QW algorithm [47–49]. Localization is destroyed in the presence of a dephasing environment which in this case acts by enhancing transport efficiency. To give an intuition on the mechanisms underlying the enhancement in ENAQT, a good example is an environment that acts on the system by causing fluctuations $\delta\varepsilon_j(t)$ in the site energies. We assume $\delta\varepsilon_j(t)$ to be independent white noise on each site, that is

$$\overline{\delta\varepsilon_j(t)} = 0, \tag{4}$$

$$\overline{\delta\varepsilon_j(t)\delta\varepsilon_{j'}(t')} = \omega_j^2 \delta_{j,j'} \delta(t-t'), \tag{5}$$

where the overbar indicates the average over the noise realizations, $\omega_j^2$ is the variance of the noise, $\delta_{j,j'}$ and $\delta(t-t')$ are respectively the Kronecker and Dirac delta functions, meaning that noise is spatially and dynamically uncorrelated. Under these assumptions, we define a stochastic time-dependent Hamiltonian as

$$H_{HS}(t) = H + H_{fluc}(t) = \sum_{j=1}^{N}\left(\varepsilon_j + \delta\varepsilon_j(t)\right)|j\rangle\langle j| + \sum_{\langle j,j' \rangle} V_{j,j'}\left(|j\rangle\langle j'| + |j'\rangle\langle j|\right). \tag{6}$$

This model was introduced by Haken and Strobl [38] to describe the effects of the phonons on the dynamics of triplet excitons in molecular crystals. When the exciton bandwidth is smaller than the phonon bandwidth, the coupling with the phonon can be modelled by Gaussian white-noise modulation of the site energies, specified by the statistical properties in eqs. (4)-(5). The fluctuations cause instantaneous and stochastic resonances of the site energies, overcoming localization and facilitating exciton transport. The state of the system acquires a random character, and the density matrix is recovered as the average over the noise realizations, $\rho(t) = \overline{|\psi(t)\rangle\langle\psi(t)|}$. The following master equation in the Lindblad form [50] is derived for the time evolution

$$\dot{\rho}(t) = -i[H,\rho(t)] + \sum_{j=1}^{N}\gamma_j\left(L_j\rho(t)L_j^\dagger - \frac{1}{2}L_j^\dagger L_j\rho(t) - \frac{1}{2}\rho(t)L_j^\dagger L_j\right), \tag{7}$$

where the Lindblad operators are simply the projectors on the site basis, $L_j = |j\rangle\langle j|$, with the resulting dissipator describing decoherence between different sites at a rate controlled by the fluctuation amplitudes, $\gamma_j = \omega_j^2$. It was shown that the simple site-dephasing dynamics as described by eq. (7) captures qualitatively the transition from a purely coherent to incoherent regimes of energy transfer as a function of the dephasing rates [51] and allows studying how the transport efficiency may be enhanced in the presence of a dephasing environment [17,18,52]. On the other hand, eq. (7) corresponds to the high-temperature limit of dephasing quantum dynamics. Indeed, the long-time asymptotic solution implies an equal occupation of all quantum sites (and all eigenstates) independently of their energy.

## 2.1 Transport efficiency

The common setting to evaluate the effect of dephasing on quantum transport is to consider the dynamical evolution of an initially localized state. In the network, a target site is identified, and the transport efficiency depends on the probability for the quantum walker to visit the target site. Often, non-Hermitian contributions to the Hamiltonian are included to model recombination of the exciton (loss of the walker) and the trapping of the exciton at the target site [53,54]. Because here we want to consider the simplest model where decoherence is the only non-unitary dynamics included in the system evolution, we define the transport efficiency like in Maier *et al.* [27], as the cumulative probability of finding the walker at the target site $j$ within a finite time $T$

$$\eta_j(T) \equiv \int_0^T p(j, t \mid j_0) dt, \qquad (8)$$

where $p(j, t \mid j_0)$ is the probability to find the walker at site $j$ at time $t$ if the dynamics started at site $j_0$ at time $t = 0$, in analogy with eq. (2). If the walker represents the exciton transport in molecular complexes, $T$ should be chosen long enough to observe ENAQT, but shorter than the typical lifetime of the exciton as recombination is not included in the dynamics.

Because the digital simulation requires to discretize the dynamics in time steps $\delta\tau$, we will approximate the integral in eq. (8) with the sum

$$\eta_j(T) \approx \sum_{s=0}^{S} p(j, s\delta\tau \mid j_0) \delta\tau, \qquad (9)$$

where $s$ is an integer indexing the time step of the simulated dynamics ranging from zero to $S$, where $S$ indicates the maximum number of time steps and $T = S\delta\tau$. Below, we will calculate the transport efficiency through different quantum algorithms by starting from an initially localized state. However, the algorithms are general, and they may be used to simulate the dynamics of any localized or diffuse, pure or mixed, initial state.

## 2.2 Mappings

The problem of simulating the quantum dynamics of a quantum molecular network as described above can be mapped in two different ways into a quantum register. We call them the *physical mapping* and the *algorithmic mapping*. Let us assume we want to simulate a chromophore network with $N$ sites. The physical mapping is inspired by the direct representation of each chromophore by a qubit of the register and it is commonly employed in analog quantum simulators [21–27]. In the simple case where each chromophore is modelled as a two-level system, the mapping on a digital quantum computer is straightforward since each chromophore corresponds to a qubit of the register. The Frenkel exciton Hamiltonian of eq. (3) is written in terms of the qubit Hamiltonian as

$$H_{ex} = -\sum_{j=1}^{N} \frac{\varepsilon_j}{2} \sigma_z^j + \sum_{\langle j,j' \rangle} \frac{V_{j,j'}}{2} \left( \sigma_x^j \sigma_x^{j'} + \sigma_y^j \sigma_y^{j'} \right), \tag{10}$$

where $\varepsilon_j$ is the energy gap between the ground and the excited state of the chromophore $j$ when it is isolated, and $\sigma_\alpha^j = \mathbb{I}^{\otimes j-1} \otimes \sigma_\alpha \otimes \mathbb{I}^{\otimes N-j}$ denotes a Pauli operator acting on the qubit in position $j$, with $\mathbb{I}$ the two-dimensional identity matrix. The populations of the states of a qubit correspond to the populations of the chromophores, and the coupling between two chromophores is realized by coupling the qubits. Within this mapping, the master equation in eq. (7) is written as

$$\dot{\rho}(t) = -i\left[H_{ex}, \rho(t)\right] + \sum_{j=1}^{N} \Gamma_j \left( \sigma_z^j \rho(t) \sigma_z^j - \rho(t) \right), \tag{11}$$

where $\Gamma_j = \gamma_j/4$, while the probability to find the walker at site $j$ is $p(j,t \mid j_0) = \langle j | \rho(t) | j \rangle$, where the basis vector identifies the state where all the qubits are in the ground state, except qubit $j$ that is excited, namely

$$|j\rangle \equiv |0\rangle_1 \otimes |0\rangle_2 \otimes ... \otimes |1\rangle_j \otimes ... \otimes |0\rangle_N \equiv |00...1_j...0\rangle. \tag{12}$$

The number of qubits required for this encoding scales linearly with the size of the system $N$ but the basis vectors which correspond to a useful state of the system are only a small subset of the entire computational basis. Although this mapping offers a vivid physical representation, it is not advantageous in terms of memory requirements. Indeed, the Hamiltonian (10) and the site-dephasing master equation (11) do not mix different exciton manifolds, and therefore to represent the single-exciton manifold a Hilbert space of dimension $N$ is sufficient, much smaller than the $2^N$ dimensional Hilbert space which is generated by the physical mapping. The exponential scaling of the computational space can be harnessed by adopting the algorithmic mapping which is usually required for a resource-efficient implementation of the quantum walk algorithm. A graph with $N$ nodes can be encoded by using only $\lceil \log_2(N) \rceil$ qubits by using the binary representation of the node index in the computational basis, that is $\left|j\right\rangle = \left|\text{bin}\left(j-1\right)\right\rangle$ for $j$ running from 1 to $N$. For example, a four-site network is efficiently encoded by two qubits as

$$\left|1\right\rangle \equiv \left|00\right\rangle \quad \left|2\right\rangle \equiv \left|01\right\rangle \quad \left|3\right\rangle \equiv \left|10\right\rangle \quad \left|4\right\rangle \equiv \left|11\right\rangle \tag{13}$$

From the perspective of a chromophore network, this mapping only represents the single-exciton manifold. Because of the exponential saving in memory obtained by encoding the position as a binary number in a quantum computer, the algorithmic mapping will always win eventually as the molecular network we want to simulate gets larger. On the other hand, because each site of the network is represented by multiple qubits, the Hamiltonian evolution requires the implementation of unitary operations over multiple qubits (eventually decomposed in a series of elementary gates) and reading the population of a single site requires measuring all the qubits. The interplay between memory advantage and circuit complexity is a key factor for estimating the scaling of the algorithmic efficiency with the size of the simulated system. The algorithms we present below can be realized by using both mappings. To illustrate both the possibilities, we will discuss explicitly the classical noise algorithm in an algorithmic mapping while we will move to a physical mapping to present the collision algorithm. In Section 5.1 the algorithmic scaling in the two cases will be analysed in detail.

## 3    Classical noise algorithm

The non-unitary time evolution described by eq. (7) physically arises from tracing out degrees of freedom which are part of the global system evolving unitarily. However, the same incoherent dynamics can also arise as a consequence of an averaging procedure over distinct autonomous evolutions (*i.e.*, trajectories). This point of view was taken by Haken and Strobl to model the effect of the phonon coupling on the excitonic transport in molecular crystals. The dephasing master equation is then obtained by averaging over the noise realizations. The first algorithm we consider is based on the simulation of the unitary dynamics ruled by the Haken-Strobl stochastic Hamiltonian in eq. (6), where the stochastic part is a classical Gaussian stochastic process specified by eqs. (4)-(5). The non-unitary evolution is then recovered by averaging over the different realizations of the circuit.

For the numerical solution of the Schrödinger equation, we first discretize the evolution time from $t=0$ to $T$ into $S$ small time intervals $\delta\tau = T/S$. A trajectory $\xi$ is obtained by evolving the initial state through the time ordered evolution operator

$$U_\xi\left(t,0\right) = \mathcal{T}\exp\left[-i\int_0^t H_{HS}\left(\tau\right)d\tau\right] \approx \prod_{s=1}^S U_\xi\left(s\delta\tau,(s-1)\delta\tau\right). \tag{14}$$

In the right-hand side, the operator is evaluated as the product of short-time evolutions governed by the average Hamiltonian in the corresponding time interval, *i.e.*,

$$U_\xi\left(s\delta\tau,(s-1)\delta\tau\right) = \exp\left[-i\int_{(s-1)\delta\tau}^{s\delta\tau} H_{HS}(\tau)d\tau\right] = \exp\left[-iH\delta\tau - i\int_{(s-1)\delta\tau}^{s\delta\tau} H_{fluc,\xi}(\tau)d\tau\right]. \quad (15)$$

The integral of the fluctuating part of the Hamiltonian requires some care because it implies integrating Gaussian white noises

$$\int_{(s-1)\delta\tau}^{s\delta\tau} H_{fluc,\xi}(\tau)d\tau = \sum_{j=1}^{N}|j\rangle\langle j|\int_{(s-1)\delta\tau}^{s\delta\tau} \delta\varepsilon_{j,\xi}(\tau)d\tau. \quad (16)$$

Each integral is a Gaussian random variable with zero mean and variance $\omega_j^2 \delta\tau$ [55]. By defining the fluctuating Hamiltonian for each time interval $s$ by a set of normally distributed energies $\delta\varepsilon_{j,\xi,s}$ with zero mean and variance $\omega_j^2$

$$H_{fluc,\xi,s} = \sum_{j=1}^{N} \delta\varepsilon_{j,\xi,s}|j\rangle\langle j|, \quad (17)$$

the short-time evolution operator (15) explicitly reads

$$U_\xi\left(s\delta\tau,(s-1)\delta\tau\right) = \exp\left[-iH\delta\tau - iH_{fluc,\xi,s}\sqrt{\delta\tau}\right]. \quad (18)$$

In Appendix A, we show that the dephasing master equation (7) can be recovered from the average of the trajectories generated by the evolution operator (14) for different realizations of the noise $\xi$ in the limit of vanishing time step $\delta\tau \to 0^+$ and for an infinite number of trajectories, *i.e.*, $\Xi \to \infty$.

Before running the algorithm, the following parameters need to be set: the time step $\delta\tau$, the number of trajectories $\Xi$, the fluctuation amplitude at each site $\omega_j^2$ and the number of samples (*shots*) that are taken during measurement. In the following, we detail the algorithmic steps to propagate the dynamics and to calculate the transport efficiency, eq. (9). Figure 1 depicts the structure of the algorithm using three qubits which can encode a network with a maximum of $2^3 = 8$ sites:

1. Initialize the quantum register to the initial state of the dynamics $|j_0\rangle$;

2. Draw $N$ independent real random numbers from a Gaussian distribution with zero mean and variance $\omega_j^2$. Each random number represents the site-energy fluctuation $\delta\varepsilon_{j,\xi,s}$;

3. Apply the unitary gate corresponding to the operator $U_\xi\left(s\delta\tau,(s-1)\delta\tau\right)$, eq. (18);

4. Create a copy of the circuit, so that one has two identical objects: a quantum circuit $\mathcal{A}$ that will be the lead and its copy $\mathcal{B}$ to which the measurement gates will be appended;

5. Add measurement gates to circuit $\mathcal{B}$ and execute it to obtain a finite-sampling estimator $\pi_\xi(j, s\delta\tau \mid j_0)$ of $p_\xi(j, s\delta\tau \mid j_0)$, which is the probability of being at the target site during the trajectory generated by $\xi$;

6. Discard circuit $\mathcal{B}$ and continue with circuit $\mathcal{A}$;

7. Repeat steps 2-6 for $s \in \{0, 1, ..., S\}$. At the end, a trajectory is embedded in the gates of circuit $\mathcal{A}$. Each time point of the trajectory is recorded in $\pi_\xi(j, s\delta\tau \mid j_0)$;

8. Repeat steps 1-7 for $\xi \in \{1, 2, ..., \Xi\}$ to obtain a swarm of trajectories;

9. Estimate the transport efficiency as

$$\eta_j(T) \approx \sum_{s=0}^{S} p(j, s\delta\tau \mid j_0)\delta\tau \approx \frac{1}{\Xi}\sum_{s=0}^{S}\sum_{\xi=1}^{\Xi} p_\xi(j, s\delta\tau \mid j_0)\delta\tau \approx \frac{1}{\Xi}\sum_{s=0}^{S}\sum_{\xi=1}^{\Xi} \pi_\xi(j, s\delta\tau \mid j_0)\delta\tau, \quad (19)$$

where each term in the equation evidences an approximation made in the estimation. Namely, the discretization of time into intervals $\delta\tau$, the finite number of trajectories $\Xi$ and the finite number of sampling shots in the final measurement of the qubits.

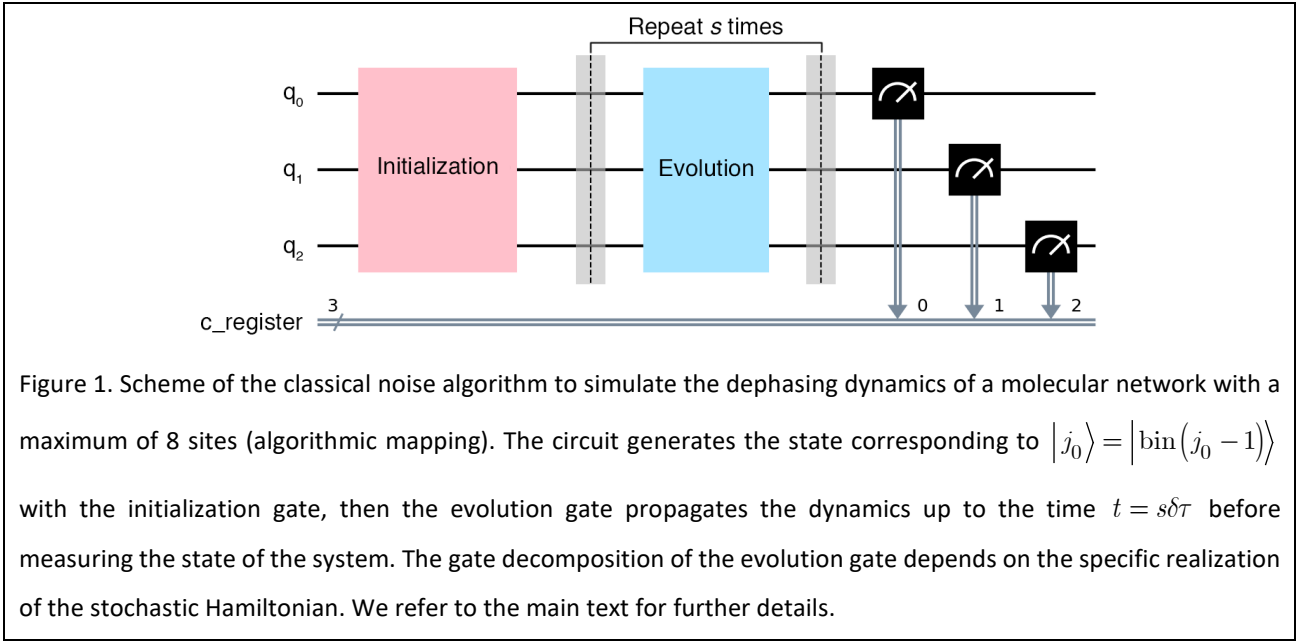

Figure 1. Scheme of the classical noise algorithm to simulate the dephasing dynamics of a molecular network with a maximum of 8 sites (algorithmic mapping). The circuit generates the state corresponding to $|j_0\rangle = |\text{bin}(j_0 - 1)\rangle$ with the initialization gate, then the evolution gate propagates the dynamics up to the time $t = s\delta\tau$ before measuring the state of the system. The gate decomposition of the evolution gate depends on the specific realization of the stochastic Hamiltonian. We refer to the main text for further details.

Notice that the initialization, evolution and measurement gates act in general on the whole quantum register as a consequence of the algorithmic mapping. Executing and measuring a clone of the circuit (step 4 of the algorithm) is important to guarantee the correct evolution of the unitary trajectory. In fact, applying measurements after every evolution gate would destroy the quantum coherence of the register, resulting in our case in a localization of the walker at some site at each time step. In step 5, a good estimator of the site population $\pi_\xi(j, s\delta\tau \mid j_0)$ is obtain by collecting a large statistical sample. However, if one is not interested in individual trajectories but only in the simulation of the master equation, it is fair to set the number of sampling shots equal to 1 to lower the execution time.

### 3.1 Stochastic trajectories and average evolution

The classical noise algorithm is implemented for the simulation of the dynamics in a $N = 4$ cyclic network with diagonal static disorder. The site energies were obtained from a Gaussian distribution centred at zero and parametrized by its standard deviation $\sigma/V = 2$, where we assume the nearest-neighbour interaction strength $V$ as the energy unit. The result is the following set of energies: $\varepsilon_1/V \simeq 0.44$, $\varepsilon_2/V \simeq 0.24$, $\varepsilon_3/V \simeq -3.22$ and $\varepsilon_4/V \simeq 0.36$, which is used for all the simulations discussed hereafter. The dynamics was initialized at the site $j_0 = 1$ and we set the target site at $j = 3$ so that we aim at computing the probability of finding the walker at the target site as a function of time. The amplitudes of the stochastic fluctuations in the Haken-Strobl Hamiltonian were set equal for every site resulting in a (scaled) dephasing rate of $\gamma_j/V = \gamma/V = 10^{-1}\hbar^{-1}$.

First, let us look at a single stochastic trajectory, this can be obtained by setting $\Xi = 1$ in the algorithm and by choosing an appropriate number of shots in step 5 to determine the average population of the target site. Figure 2a shows such a stochastic trajectory (blue line) compared with the dynamics of the isolated system (green line), that is in the absence of white noise. Note how the single trajectory evolves differently from the isolated system from the very beginning of the dynamics. However, as for the isolated system, the single trajectory never reaches an equilibrium constant value. A swarm of $\Xi = 200$ unitary stochastic trajectories obtained from different realizations of the noise is reported in Figure 2b (blue lines in transparency). The trajectory of the open system is then obtained as an average of them (solid blue), and it is in very good agreement with the direct solution of the master equation (orange dotted line). The decay of the site coherences and the emergence of a constant equilibrium value for the site population are not a direct consequence of the presence of the noise, but rather emerge because of the averaging procedure.

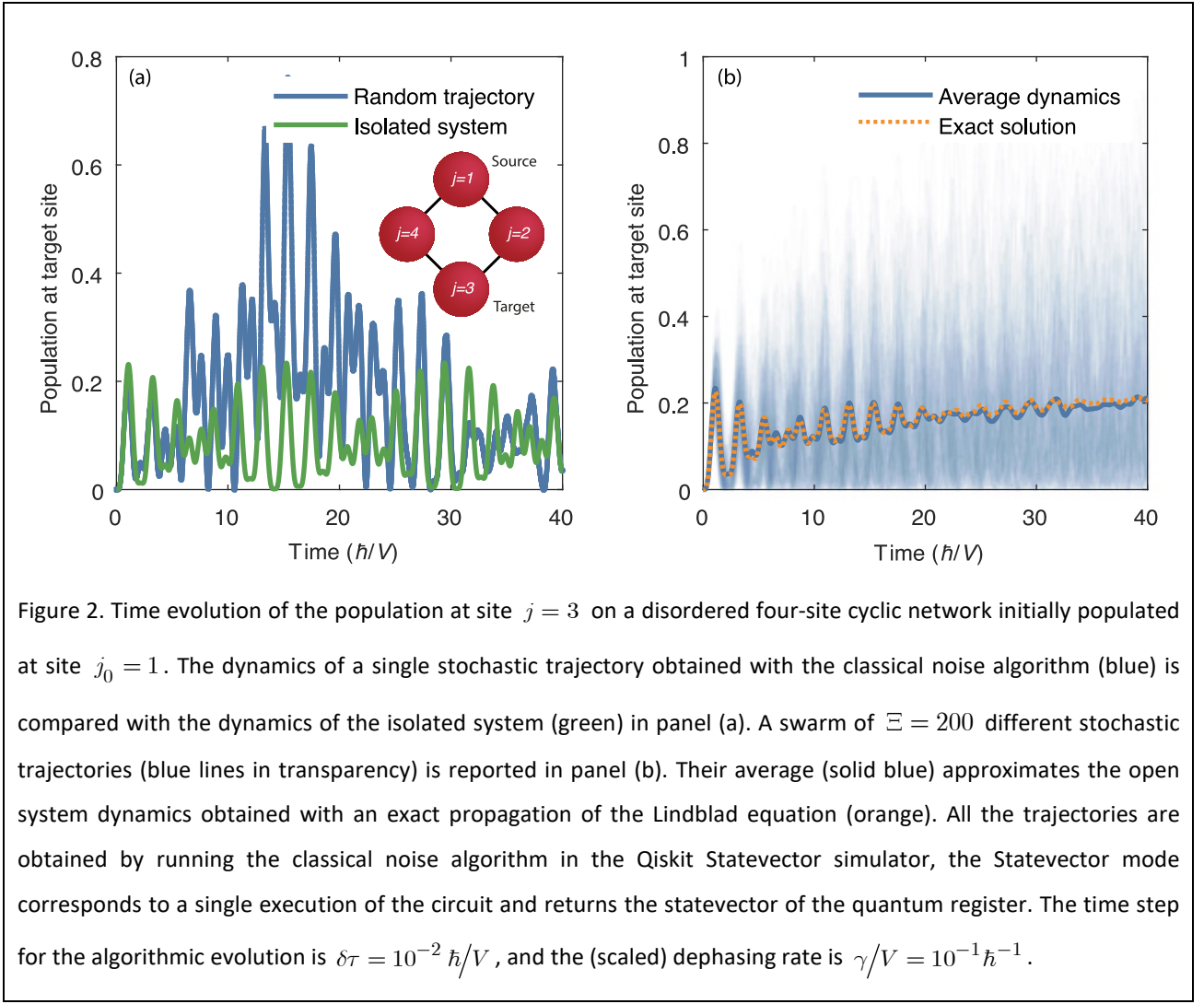

Figure 2. Time evolution of the population at site $j = 3$ on a disordered four-site cyclic network initially populated at site $j_0 = 1$. The dynamics of a single stochastic trajectory obtained with the classical noise algorithm (blue) is compared with the dynamics of the isolated system (green) in panel (a). A swarm of $\Xi = 200$ different stochastic trajectories (blue lines in transparency) is reported in panel (b). Their average (solid blue) approximates the open system dynamics obtained with an exact propagation of the Lindblad equation (orange). All the trajectories are obtained by running the classical noise algorithm in the Qiskit Statevector simulator, the Statevector mode corresponds to a single execution of the circuit and returns the statevector of the quantum register. The time step for the algorithmic evolution is $\delta\tau = 10^{-2}\,\hbar/V$, and the (scaled) dephasing rate is $\gamma/V = 10^{-1}\hbar^{-1}$.

The classical noise algorithm as described above can be readily adapted to other models of stochastic modulation of the Hamiltonian (coloured noise) to investigate the role of the environment correlation time in the decoherence dynamics and energy transfer rates in molecular systems. In this case, the average dynamics is not known *a priori* because stochastic modulations different from white noise do not lead in general to a Lindblad form of the master equation [56–58]. A simple choice to investigate these effects, as suggested for example by Jansen *et al* [59], is to assume that the energy modulations in the Hamiltonian eq. (6) are described by Ornstein-Uhlenbeck processes with temporal correlation $\overline{\delta\varepsilon_j(t)\delta\varepsilon_{j'}(t')} = \omega_j^2 \delta_{j,j'} \exp(-\Lambda|t-t'|)$, with $\Lambda^{-1}$ the correlation time. A realization of the fluctuation process can be generated at each time step $s$ as $\delta\varepsilon_{j,\xi,s} = \delta\varepsilon_{j,\xi,s-1}\exp(-\Lambda\delta\tau) + G_{j,\xi,s}\sqrt{1-\exp(-2\Lambda\delta\tau)}$, where $G_{j,\xi,s}$ is a normally distributed random number with zero mean and variance $\omega_j^2$ and the process is initialized as $\delta\varepsilon_{j,\xi,s=0} = G_{j,\xi,s=0}$ [60]. Once the evolution operators $U_\xi(s\delta\tau,(s-1)\delta\tau)$ in eq. (14) are built accordingly, the structure of the algorithm (Figure 1) does not change.

## 4 Collision algorithm

Quantum collision models have recently emerged as a powerful and versatile tool to describe the dynamics of open quantum systems. The basic idea is to induce decoherence and relaxation in the dynamics of an open system by repeated interactions (*collisions*) between the system and a set of ancillae that represents an effective environment. Currently, collision models are used in the study of quantum thermodynamics [61–63], quantum non-Markovian dynamics [64–66], quantum optics [67] and foundational issues such as quantum Darwinism [68]. Ref. [69] gives a perspective on the application of collision models in quantum physics. Recently, a stochastic version of a collision model has been applied to the study of ENAQT [70]. Beside being widely used both for theoretical investigations and simulations on classical computers, collision models are particularly suitable to be translated in quantum algorithms. Literature in this sense is very recent and limited to single-qubit dynamics [71,72]. In this Section, we discuss a quantum algorithm for the simulation of dephasing-enhanced quantum transport based on collisions between the system and the ancillae representing the environment.

To introduce the underlying idea, imagine we associate at each site of the molecular network a reservoir of ancilla systems. As in the classical noise algorithm, we first discretize the evolution time. During each time step $\delta\tau$ (collision time), each site interacts with one ancilla of its reservoir. The dynamics proceeds through successive pairwise collisions between the system and the bath ancillae, each collision involving a unitary evolution of the system and the colliding ancilla. If the ancillae are initially uncorrelated and each of them collides with the system only once, then it can be shown that the dynamics of the system is described by the solution of a Lindblad master equation in the continuous-time limit [39,73].

The concept of a reservoir of independent ancillae is used in the framework of collision models and we will use it to give the theoretical basis of the algorithm. However, we will show later that for the algorithm a single ancilla-qubit is sufficient to represent the whole environment.

We will work in a physical mapping of the system, where each qubit represents a site of the network and the system Hamiltonian is translated in the qubit Hamiltonian of eq. (10). During a time interval $\delta\tau$, we define the Hamiltonian of the collision model acting on the system-ancillae state as

$$H_{CM} = H_{ex} \otimes \mathbb{I}^{\otimes N} + H_{int} =$$
$$= \sum_{j=1}^{N} \frac{\varepsilon_j}{2} \sigma_z^j \otimes \mathbb{I}^{\otimes N} + \sum_{\langle j,j'\rangle} \frac{V_{j,j'}}{2} \left( \sigma_x^j \sigma_x^{j'} \otimes \mathbb{I}^{\otimes N} + \sigma_y^j \sigma_y^{j'} \otimes \mathbb{I}^{\otimes N} \right) + \sum_{j=1}^{N} c_j \sigma_z^j \otimes \sigma_x^{a_j} \quad (20)$$

where $H_{int}$ is the interaction Hamiltonian between the sites and the corresponding ancillae and $\sigma_x^{a_j}$ is the Pauli X-operator acting on the environment ancilla assigned to site $j$. We emphasise the separation between the system $\mathcal{S}$ and the ancillary $a$ subspaces by the explicit tensor product $\otimes$. The choice of the interaction Hamiltonian is crucial to recover a specific dynamics of the system $\mathcal{S}$. By assuming a $\sigma_z^j \otimes \sigma_x^{a_j}$ form of the interaction, we obtain the dephasing dynamics described by eq. (11) when the initial state of the ancilla system is any state which is diagonal in the computational basis (see eq. (34) in Appendix B). For convenience, we set the initial state of each ancilla as the qubits' ground state, $\rho_a = \left(|0\rangle\langle 0|\right)^{\otimes N}$. After the interaction, the ancillae are discarded and no longer affect the dynamics of the system, therefore we trace over their degrees of freedom to obtain the reduced density matrix of the system. Thus, the dynamical map of a collision event can be written as

$$\rho_\mathcal{S}(t+\delta\tau) = \Phi[\rho_\mathcal{S}(t)] = \text{Tr}_a\left\{U\left(\rho_\mathcal{S}(t)\otimes\rho_a\right)U^\dagger\right\}, \qquad (21)$$

where $\text{Tr}_a$ is the trace over the ancillary degrees of freedom and $U = \exp(-iH_{CM}\delta\tau)$ is the evolution operator. By sequentially repeating the interaction step with new ancillae from the reservoir, the reduced dynamics of the system at later times is obtained

$$\rho_\mathcal{S}(s\delta\tau) = \Phi_s[\rho_\mathcal{S}(0)] = \Phi\left[\Phi\left[...\Phi\left[\rho_\mathcal{S}(0)\right]...\right]\right]. \qquad (22)$$

In Appendix B we show that, in the limit $\delta\tau \to 0^+$, the dynamical map of eq. (22) approximates the solution of the target master equation (11) with dephasing rates $\Gamma_j = c_j^2 \delta\tau$. The dephasing rates are therefore determined by the strength of the interaction and the collision time, conveying the intuition that the system undergoes decoherence when it interacts strongly for a short time as well as when the interaction is weaker, but it lasts longer. Note that the dependence of the interaction strength from the collision time $c_j = \sqrt{\Gamma_j/\delta\tau}$ is a typical relation in quantum collision models [67,74].

We are now in the position of translating the collision model described above into an algorithm that simulates the site-dephasing master equation. Given the decoherence rate, the user must choose the time step $\delta\tau$ and the coupling strengths with the ancillae $c_j$ accordingly. Let us first translate the model into an algorithm that uses two quantum registers of the same size to represent the system and the ancillae in a physical mapping; keeping in mind that we will then shrink the register of the ancillae to a single qubit. The algorithm implements the following steps, graphically represented in Figure 3a:

1. Initialize the system quantum register to the initial state of the dynamics $|j_0\rangle$ (*i.e.*, apply an X-gate to the qubit representing the initial site of the dynamics in the physical mapping);
2. Apply the gates corresponding to the evolution operator $U$ of the collision model Hamiltonian eq. (20) to both the system and environment quantum registers;
3. Reset the environment quantum register;
4. Create a copy of the circuit, so that you have two identical objects: a quantum circuit $\mathcal{A}$ that will be the lead and its copy $\mathcal{B}$ to which the measurement gate will be appended;

5. Add a measurement gate to circuit $\mathcal{B}$, in correspondence of the qubit encoding the target site and read the population of state $1$ to obtain the estimator $\pi(j, s\delta\tau \mid j_0)$ of $p(j, s\delta\tau \mid j_0)$;

6. Discard circuit $\mathcal{B}$ and continue with circuit $\mathcal{A}$;

7. Repeat steps 2-6 for $s \in \{0, 1, ..., S\}$;

8. Approximate the transport efficiency as $\eta_j(T) \approx \sum_{s=0}^{S} p(j, s\delta\tau \mid j_0) \delta\tau \approx \sum_{s=0}^{S} \pi(j, s\delta\tau \mid j_0) \delta\tau$.

The evolution operator in step 2 which identifies the evolution gate in Figure 3a needs to be decomposed into elementary (one and two-qubit) gates. A generic unitary operation on $n$ qubits may imply an impractical exponential scaling of the length of the circuit. However, the physical mapping suggests the implementation of a Trotter-Suzuki decomposition which at the first order in the interaction reads

$$\begin{aligned} U &= e^{-i\left(\sum_{j=1}^{N} \frac{\varepsilon_j}{2}\sigma_z^j \otimes \mathbb{I}^{\otimes N} + \sum_{\langle j,j'\rangle} \frac{V_{j,j'}}{2}\left(\sigma_x^j \sigma_x^{j'} \otimes \mathbb{I}^{\otimes N} + \sigma_y^j \sigma_y^{j'} \otimes \mathbb{I}^{\otimes N}\right) + \sum_{j=1}^{N} c_j \sigma_z^j \otimes \sigma_x^{a_j}\right)\delta\tau} \\ &\approx \left(e^{-i\frac{\delta\tau}{m}\sum_{j=1}^{N} c_j \sigma_z^j \otimes \sigma_x^{a_j}} e^{-i\frac{\delta\tau}{m}\sum_{\langle j,j'\rangle} \frac{V_{j,j'}}{2}\left(\sigma_x^j \sigma_x^{j'} \otimes \mathbb{I}^{\otimes N} + \sigma_y^j \sigma_y^{j'} \otimes \mathbb{I}^{\otimes N}\right)} e^{-i\frac{\delta\tau}{m}\sum_{j=1}^{N} \frac{\varepsilon_j}{2}\sigma_z^j \otimes \mathbb{I}^{\otimes N}}\right)^m + O\left(\frac{\delta\tau^2}{m^2}\right) \\ &= \left(\prod_{j=1}^{N} e^{-i\frac{\delta\tau}{m} c_j \sigma_z^j \sigma_x^{a_j}} \prod_{\langle j,j'\rangle} e^{-i\frac{\delta\tau}{m} \frac{V_{j,j'}}{2}\left(\sigma_x^j \sigma_x^{j'} + \sigma_y^j \sigma_y^{j'}\right)} \prod_{j=1}^{N} e^{-i\frac{\delta\tau}{m} \frac{\varepsilon_j}{2}\sigma_z^j}\right)^m + O\left(\frac{\delta\tau^2}{m^2}\right) \end{aligned} \quad (23)$$

with $m \in \mathbb{N}$. The unitary operator is eventually approximated by the sequential application of one and two-qubit operators, where the approximation improves for large $m$. We note that this form of the evolution operator is particularly convenient because each site-ancilla interaction occurs as a standalone process and therefore one ancilla qubit which is reset after each interaction suffices to implement the whole environment (as represented in Figure 4). Moreover, as the collision model is shown to converge to the solution of the Lindblad master equation in the limit of small collision time $\delta\tau$, a suitable choice of the time step can be made to ensure the approximation of eq. (23) is already satisfactory for $m = 1$. The effects of the time step will be discussed later in Section 5. The application of the Trotter-Suzuki decomposition allows us to further specify the circuit of the evolution gate in step 2 by reducing the ancilla system to a single qubit, as follow:

2a. Apply the $\mathrm{RZ}\left(-\varepsilon_j \delta\tau/m\right)$ gates to all the qubits of the system quantum register;

2b. Apply the $\mathrm{RXX}\left(V_{j,j'}\delta\tau/m\right)$ and $\mathrm{RYY}\left(V_{j,j'}\delta\tau/m\right)$ gates to all the system qubits representing coupled sites of the network;

2c. Apply the $\mathrm{RZX}\left(2c_j \delta\tau/m\right)$ to the $j$-th system qubit and the ancilla;

2d. Reset the state of the ancilla;

2e. Repeat steps 2c-2d for $j \in \{1, 2, ..., N\}$;

2f. Repeat steps 2a-2e $m$ times.

The circuit which realizes the evolution operations by employing a single ancilla qubit is reported in Figure 3b for $m=1$. As mentioned before, the collision algorithm can be applied also when the system is encoded in the quantum computer through an algorithmic mapping. In that case, the most convenient form of the interaction Hamiltonian and the state of the ancilla are different and are discussed in Appendix C.

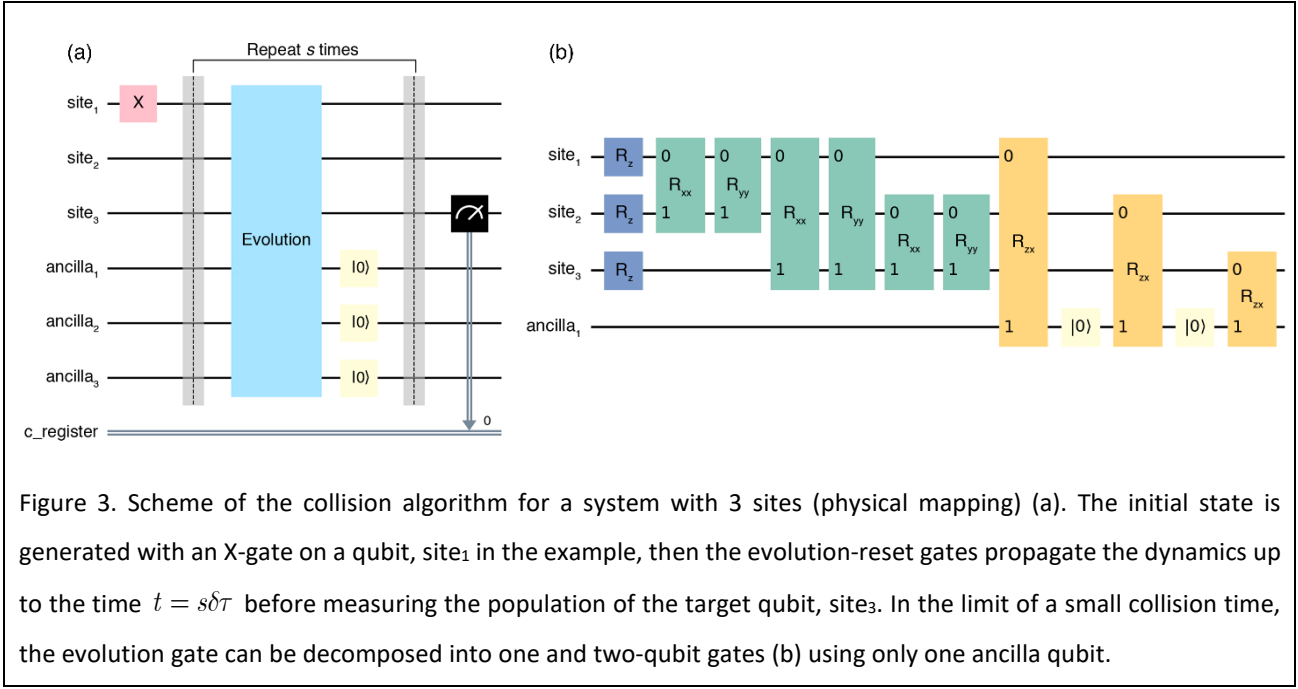

Figure 3. Scheme of the collision algorithm for a system with 3 sites (physical mapping) (a). The initial state is generated with an X-gate on a qubit, site$_1$ in the example, then the evolution-reset gates propagate the dynamics up to the time $t = s\delta\tau$ before measuring the population of the target qubit, site$_3$. In the limit of a small collision time, the evolution gate can be decomposed into one and two-qubit gates (b) using only one ancilla qubit.

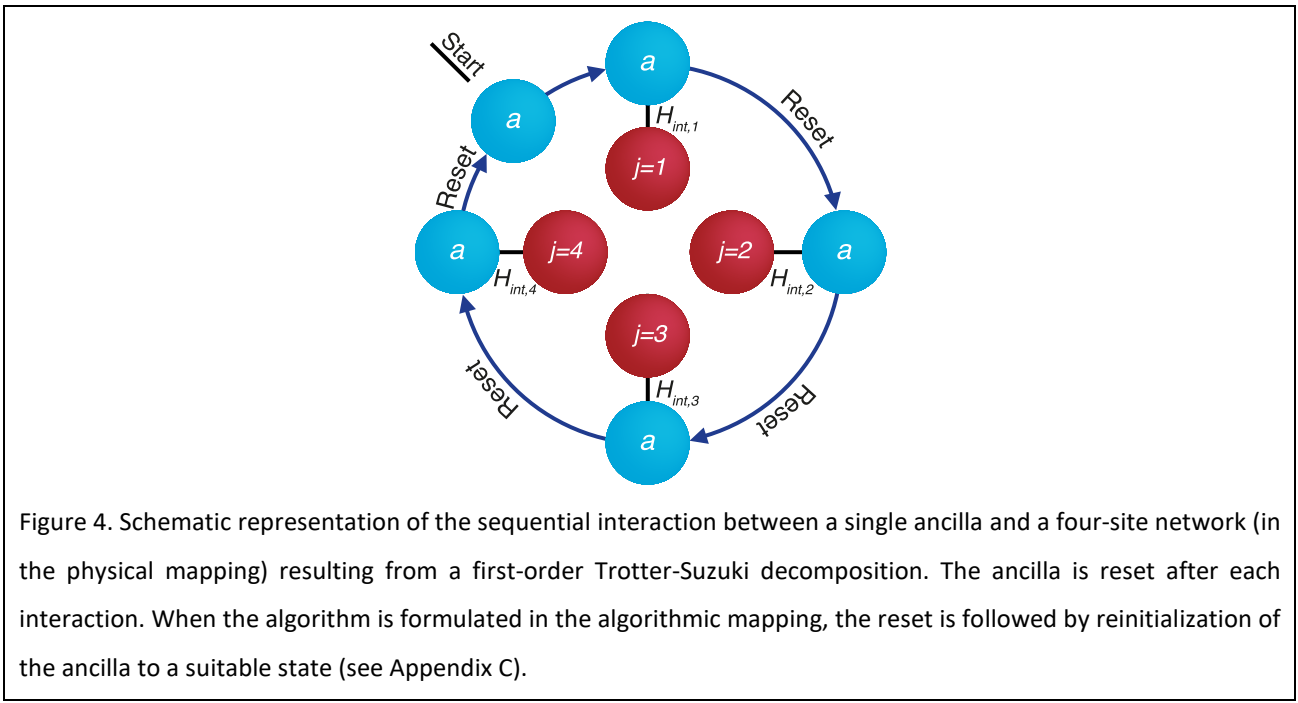

Figure 4. Schematic representation of the sequential interaction between a single ancilla and a four-site network (in the physical mapping) resulting from a first-order Trotter-Suzuki decomposition. The ancilla is reset after each interaction. When the algorithm is formulated in the algorithmic mapping, the reset is followed by reinitialization of the ancilla to a suitable state (see Appendix C).

## 5   Comparative analysis of the two algorithms

We now test both the classical noise algorithm and the collision algorithm by simulating the open system dynamics of the disordered four-site network already introduced in Section 3.1. The implementation of the classical noise algorithm in the algorithmic mapping requires two qubits, which are sufficient to encode the four states of the network. In contrast, the collision algorithm is run on five qubits, four qubits to encode the network in the physical mapping and one ancilla qubit encoding the dephasing environment. By comparing the circuits shown in Figure 1 (classical noise algorithm) and Figure 3 (collision algorithm), first note the differences imposed by the different mappings. The algorithmic mapping requires the preparation of the initial state (initialization of the circuit) and the readout of the population of the target site to be implemented by multiple qubit gates. In the physical mapping the same operations are performed by simpler single-qubit gates. Starting from site $j_0 = 1$, the population of the target site $j = 3$ is calculated as a function of time and the results are shown in Figure 5 together with the numerical solution of the Lindblad dynamics for the set of parameters specified in the legend. For the classical noise algorithm (Figure 5a) we generated $\Xi = 8000$ trajectories, measuring each one with a single shot for each time point of the dynamics. The results for the collision algorithm (Figure 5b) were obtained with 8000 shots from the same circuit for each time point. For both algorithms we used a time step $\delta\tau = 10^{-2}\,\hbar/V$. The algorithms were executed on the noiseless Qiskit QASM simulator.

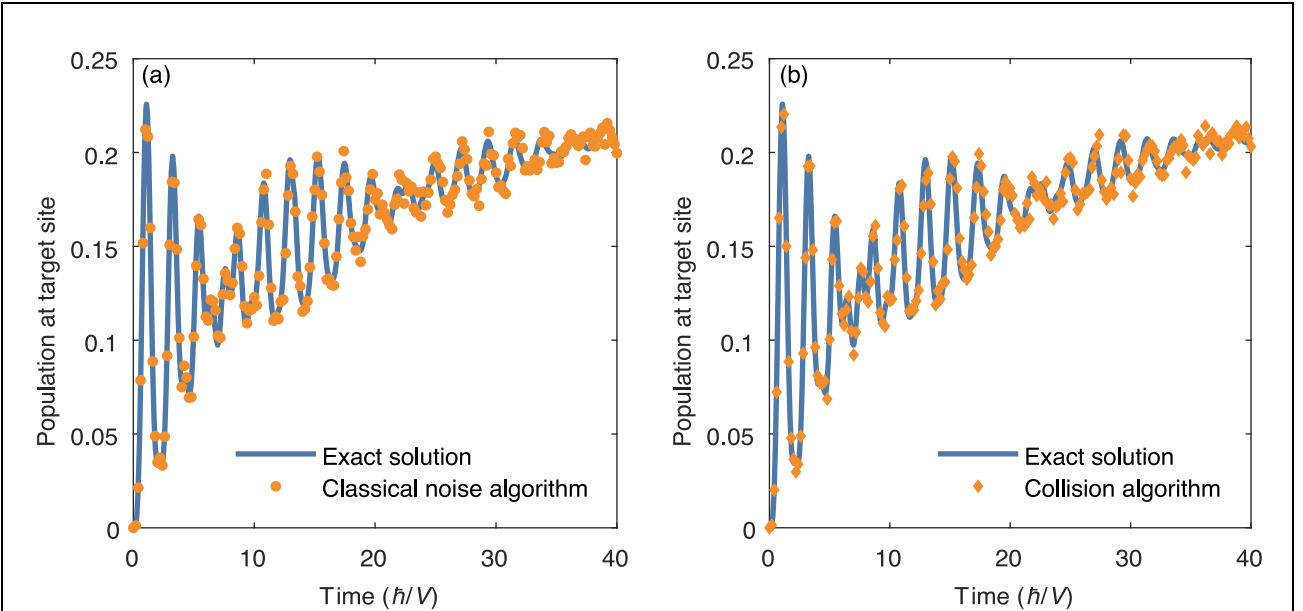

Figure 5. Time evolution of the population at site $j = 3$ on a disordered four-site cyclic network initially populated at site $j_0 = 1$. Exact numerical solution (blue line) refers to the solution of the Lindblad master equation. Results for the classical noise algorithm (a) and collision algorithms (b) are acquired by running the algorithms with the Qiskit QASM simulator. The time step for the algorithmic evolution is $\delta\tau = 10^{-2}\,\hbar/V$, and the (scaled) dephasing rate is $\gamma/V = 10^{-1}\hbar^{-1}$.

By simulating the dynamics changing the decoherent environment, we then recover the phenomenology of ENAQT in terms of transport efficiency. Figure 6 reports the transport efficiency $\eta_{j=3}\left(T = 40\,\hbar/V\right)$ obtained from the quantum simulations as a function of the ratio between the decoherence rate and the site coupling, $\gamma/V$. Simulations were performed under the same conditions already described in Figure 5, only tuning the fluctuation amplitudes in the classical noise algorithm and the strength of the site-ancilla interaction in the collision algorithm. The results are in very good agreement with the curve obtained by solving the Lindblad master equation (blue line). The transport efficiency as a function of the decoherence rate presents a maximum in the intermediate regime of coupling with the environment, which is the hallmark of ENAQT. Low transport efficiency in the weak decoherence limit can be understood as the prevailing of the static disorder, indeed the average population on the target site would be very low in the absence of decoherence. The walker tends to be trapped in the initial site and the efficiency drops also in the strong decoherence limit, although the mechanism is different. In this case, the coherent transfer from site to site (induced by the system Hamiltonian) is suppressed by the fast decay of the site coherences. In the limit of an infinite decoherence rate, the walker would never leave the initial site and the efficiency would be zero, a limit that is often associated with the quantum Zeno paradox [75].

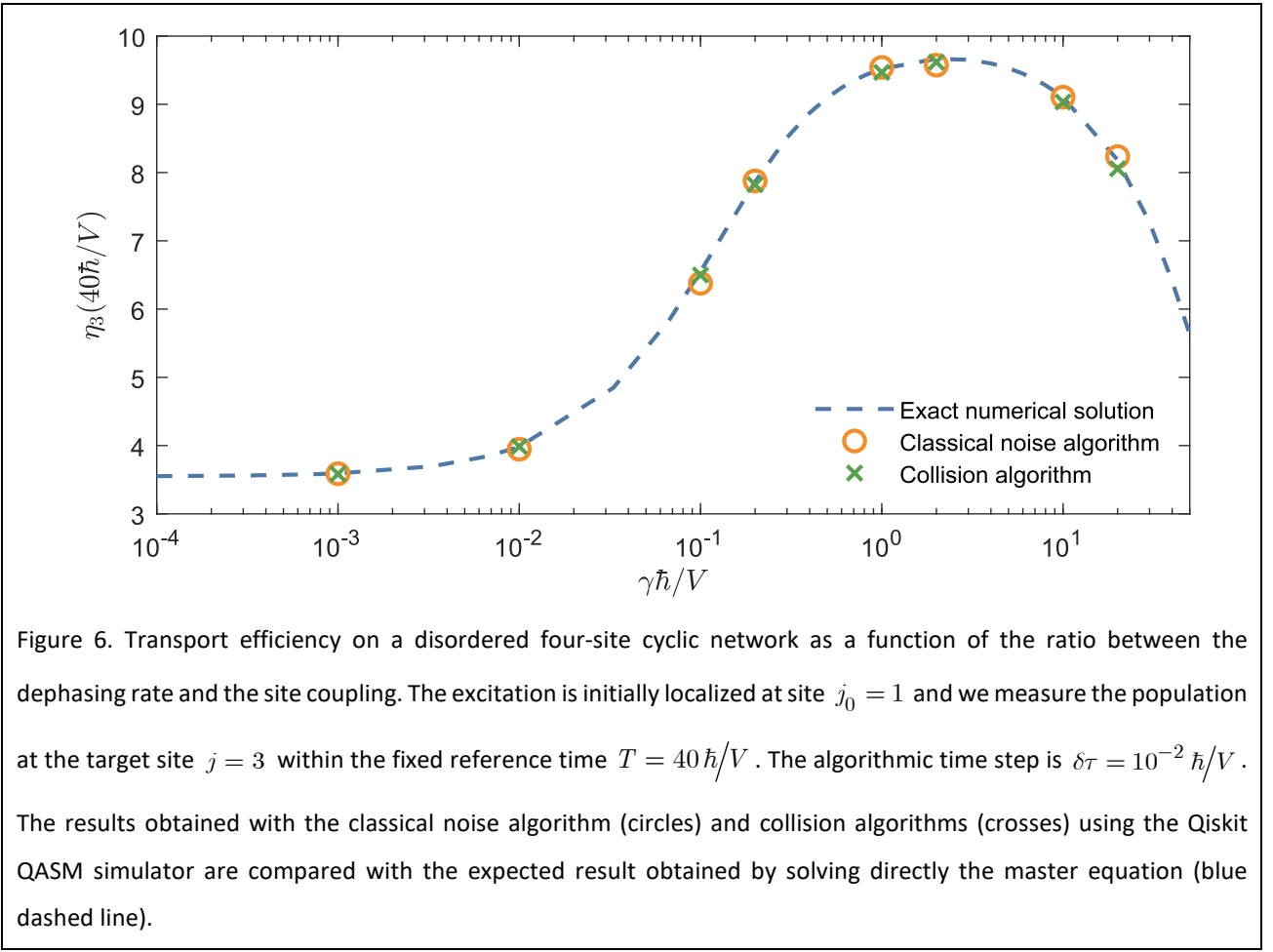

Figure 6. Transport efficiency on a disordered four-site cyclic network as a function of the ratio between the dephasing rate and the site coupling. The excitation is initially localized at site $j_0 = 1$ and we measure the population at the target site $j = 3$ within the fixed reference time $T = 40\,\hbar/V$. The algorithmic time step is $\delta\tau = 10^{-2}\,\hbar/V$. The results obtained with the classical noise algorithm (circles) and collision algorithms (crosses) using the Qiskit QASM simulator are compared with the expected result obtained by solving directly the master equation (blue dashed line).

The accuracy of the simulations performed with both the algorithms critically depends on the time step $\delta\tau$ chosen to discretize the continuous dynamics. As the time interval decreases, the simulated dynamics approximates the exact solution of the master equation with increasing accuracy (cf. Appendix A and B). On the other hand, the execution time of the circuit grows significantly and a reasonable trade-off between accuracy and simulation time must be found also considering the specific parameters of the simulated system (Hamiltonian spectrum, decoherence rate and simulation length $T$). For example, it is worth noting that as the dephasing rate increases, smaller time steps are required to maintain accuracy of the simulated dynamics. In general, a reasonable guess is to use a $\delta\tau$ which is at least one order of magnitude smaller than any characteristic timescale of the system dynamics. Operationally, the obtained result can be validated by checking that the observable of interest remains practically unchanged by halving the time step.

Another parameter to consider is the number of circuit repetitions, which assume a different meaning in the two algorithms. For the classical noise algorithm to reproduce the dynamics of the master equation, the statistics must be accumulated from different trajectories. This means that to accumulate a given number of shots, *e.g.* 8000 shots as used for the dynamics of Figure 5 and Figure 6, one needs to generate the same number of quantum circuits and perform a single-shot measurement on each of them. Each circuit is composed of different gates depending on the specific sequence of random numbers characterizing the noise of the specific trajectory. In contrast, with the collision algorithm, the same circuit is executed and measured 8000 times (*i.e.*, 8000 shots). Although this difference in the statistic accumulation of the two algorithms does not affect the analysis of the scaling of the resources presented in the next Section, one should consider that the generation of multiple circuits implies a (classical) computational overhead in the *transpilation* procedure, which is the decomposition of the circuit in elementary gates.

### 5.1 Scaling of memory and algorithmic resources

In this Section, we will analyse how the number of qubits and gates required by the algorithms scale with the size of the simulated system $N$ when using the algorithmic and the physical mappings. The results are summarized in Table 1. As already reported above, the algorithmic scaling implies an exponential advantage in terms of memory resources as the number of qubits required to encode the system scales only logarithmically with its size, $n = \lceil \log_2 N \rceil$. However, this mapping complicates the gate decomposition of the simulation procedure. The classical noise algorithm in the formulation of an algorithmic mapping consists of two operations that potentially act on the whole quantum register of size $n$: the initialization and the evolution (see Figure 1). If the initial state is completely localized on one site, the initialization process translates into a series of X-gates, whose impact on the scaling analysis is negligible. This is not the case for the evolution gate. Consisting in a unitary transformation acting on $\lceil \log_2 N \rceil$ qubits, in the general case it requires a number of CNOT gates that scales exponentially with the number of qubits of the register, and therefore quadratically with the size of the network: $O\left(4^{\lceil \log_2 N \rceil}\right) \approx O\left(N^2\right)$. Since the dynamics is obtained by a series of different evolution gates (because the system is subject to random fluctuations), each gate requires a dedicated decomposition. Although this does not compromise quadratic scaling, it involves a classical computational effort in transpilation which adds up to the effort already required to simulate different trajectories. This overhead can be slightly mitigated by adopting a suitably small time step for the dynamics. Indeed, in the limit of a vanishing time step we can assume

$$\lim_{\delta\tau \to 0^+} e^{-i\left(H\delta\tau + H_{fluc,s}\sqrt{\delta\tau}\right)} \approx e^{-iH\delta\tau} e^{-iH_{fluc,s}\sqrt{\delta\tau}}. \tag{24}$$

The first term on the right-hand side is common for the evolution at each time step and for each trajectory and therefore it can be decomposed once for all. The second term, which changes in each evolution operator, is a diagonal unitary, therefore admitting a more performant gate decomposition scaling linearly with the size of the problem, which is $O\left(2^{\lceil \log_2 N \rceil}\right) \approx O\left(N\right)$ gates. Overall, the leading term in the scaling remains quadratic. The analysis of the collision algorithm in the algorithmic mapping is reported in Appendix C. It turns out that the depth of the circuit grows also quadratically with the size of the system, while it requires $\lceil \log_2 N \rceil$ qubits and an extra ancilla qubit to represent the environment.

The physical mapping, on the other hand, is not memory efficient as it requires a number of qubits that grows linearly with the size of the network, that is $N$ for the classical noise algorithm and $N+1$ for the collision algorithm. On the other hand, it greatly simplifies the gate decomposition procedure. As we have shown in Section 4, it is possible to perform the initialization of the circuit by the application of a single X-gate, while the evolution gate, which is repeated during the algorithm, can be easily reduced to 1- and 2-qubits gates. In this case, the interaction Hamiltonian between adjacent sites gives the leading term for the scaling of the algorithm: in a network where all sites are coupled, $O\left(N^2\right)$ two-qubit gates are required to account for the unitary evolution. When the connectivity is restricted (as in tight-binding graphs), the resulting scaling is usually sub-quadratic (*e.g.*, linear for a cyclic topology).

Although the scaling with the size of the network results similar for the two mappings, the classical effort required by the algorithmic mapping for gate decomposition partially contrasts the stunning advantage in terms of memory. This "classical" overhead is further amplified in the classical noise algorithm by the need of creating a different circuit for each time step of each trajectory.

Let us now compare the scaling of the algorithms discussed above with the other quantum algorithms discussed in literature for the simulation of environment-assisted quantum transport. We find a slightly more favourable scaling than the algorithmic strategy proposed by Gupta and Chandrashekar in ref. [35], which provides a scaling of $O\left(N^2 \log_2 N\right)$ by employing an algorithmic mapping with some extra ancillae. On the other hand, the model of ref. [35] allows the introduction of the effect of the temperature at the cost of working in the exciton basis, thus requiring a preparatory step involving the diagonalization of the Hamiltonian. The algorithm discussed by Hu *et al.* in ref. [36], which is based on a dilation approach, is characterized by a remarkable sub-polynomial scaling of $O\left(\log_2^2 N\right)$. However, the authors also discuss an exponential scaling with the number of time steps of the dynamics which can be probably reduced by circuit optimization in the simulations of specific systems.

Table 1. Comparative analysis of the scaling of the memory resources and circuit depth for the classical noise and the collision algorithm with the size $N$ of the simulated system in the two different mappings. The first term refers to the number of qubits needed for the simulation while the second term refers to the gate count.

|  | Classical noise algorithm | Collision algorithm |
| --- | --- | --- |
| Physical mapping | $N$, $O\left(N^2\right)$ | $N+1$, $O\left(N^2\right)$ |
| Algorithmic mapping | $\lceil \log_2 N \rceil$, $O\left(N^2\right)$ | $\lceil \log_2 N \rceil + 1$, $O\left(N^2\right)$ |

# 6   Algorithmic quantum trajectories

*Unravelling* the master equation for the system density matrix amounts to considering an ensemble of pure states whose average gives the desired dynamics. Each system, described by a wavefunction, evolves stochastically in time defining a so-called quantum trajectory [76]. Unravelling approaches originate from the quantum optics community [77,78], when technological advancements allowed for monitoring individual quantum systems. Quantum trajectories were proved to be tightly connected with the measurement protocol used to monitor quantum states [79] providing an effective framework to describe different detection schemes, as photon-counting and homodyne detection, for which ensemble descriptions based on the density matrix does not offer a straightforward interpretation. Subsequently, quantum trajectories have been used in many other fields as a numerical technique to solve the open system dynamics with the number of variables being linear with the dimension of the Hilbert space rather than quadratic as the dimension of the density matrix [80–83]. In this section, we want to characterize the *algorithmic quantum trajectories*, which are quantum trajectories generated by a given algorithm in a quantum computer as the result of the different operations on the computer state, both in terms of gates (classical noise algorithm) and measurements on the ancillae (collision algorithm).

The very working principle of digital quantum simulations requires the accumulation of measurement statistics to evaluate the expectation value of interest, which for our purpose is the population of the target site at different times, *i.e.*, $p(j, t \mid j_0) = \langle j | \rho(t) | j \rangle$. This implies that the quantum circuits must be executed many times and the simulated open system dynamics can be naturally understood as the average of different realizations of underlying stochastic dynamics. This observation implies that different quantum algorithms realize different unravellings of the master equation.

The classical noise algorithm is based on an explicit average over trajectories which are generated by the stochastic terms in the Hamiltonian which are mapped into the gates of the quantum circuit. When the algorithmic time step is small enough each trajectory is a good approximation of a stochastic Schrödinger equation where the stochastic part is given by a Wiener increment [55]

$$d|\psi(t)\rangle = \left(-iH - \frac{1}{2}\sum_{j=1}^{N}\omega_j^2 L_j^\dagger L_j\right)|\psi(t)\rangle dt - i\sum_{j=1}^{N}\omega_j L_j |\psi(t)\rangle dW_j(t),$$

where $L_j = |j\rangle\langle j|$, $\overline{dW_j(t)} = 0$ and $\overline{dW_j(t)^2} = dt$. The stochastic differential introduces small but frequent deviations from the evolution of the isolated system ruled by the average Hamiltonian, resulting in the noisy trajectories already shown in Figure 2.

Trajectories obtained by the collision model algorithm are qualitatively different. In this case, the stochastic nature of the evolution can be more easily understood as resulting from the quantum probabilities for the outcomes of measurements performed on the ancilla environment. Indeed, at each time step after the interaction with the system, the state of the ancilla is reset. The reset gate acts as a measurement of the ancilla qubit and its subsequent reinitialization to state $|0_a\rangle$. One thus goes from an entangled system-ancilla state $|\psi\rangle = c_0 |\psi_{0,\mathcal{S}}\rangle \otimes |0_a\rangle + c_1 |\psi_{1,\mathcal{S}}\rangle \otimes |1_a\rangle$ to a product state that is $|\psi_{0,\mathcal{S}}\rangle \otimes |0_a\rangle$ with probability $|c_0|^2$ or $|\psi_{1,\mathcal{S}}\rangle \otimes |0_a\rangle$ with probability $|c_1|^2$. Performing this operation multiple times reproduce the effect of tracing out the ancilla's degrees of freedom, resulting in a statistical density matrix for the system $\rho_\mathcal{S} = |c_0|^2 |\psi_{0,\mathcal{S}}\rangle\langle\psi_{0,\mathcal{S}}| + |c_1|^2 |\psi_{1,\mathcal{S}}\rangle\langle\psi_{1,\mathcal{S}}|$. However, a single run of the circuit of the collision algorithm realizes a single quantum trajectory. The specific character of the trajectory depends on the form of the interaction between the system and the ancilla environment. Let us look closer into the algorithm steps discussed in Section 4 and reported in Figure 3: suppose we have initialized the state and executed the first block of RZ, RXX and RYY in the system register. After this evolution block, the state of the quantum register will be generically $|\psi\rangle = |\psi_\mathcal{S}\rangle \otimes |0_a\rangle$. The interaction between the first qubit of the system, $j=1$, and the ancilla is through an RZX gate, which is

$$\exp\left(-ic_1\sigma_z^1 \otimes \sigma_x^a \delta\tau\right) = \cos\left(c_1\delta\tau\right)\mathbb{I}^1 \otimes \mathbb{I}^a - i\sin\left(c_1\delta\tau\right)\sigma_z^1 \otimes \sigma_x^a. \quad (25)$$

By applying it to the state $|\psi\rangle$ we explicitly get

$$\begin{aligned}|\psi'\rangle &= \cos\left(c_1\delta\tau\right)\mathbb{I}^1 \otimes \mathbb{I}^a |\psi\rangle - i\sin\left(c_1\delta\tau\right)\sigma_z^1 \otimes \sigma_x^a |\psi\rangle \\ &= \cos\left(c_1\delta\tau\right)|\psi_\mathcal{S}\rangle \otimes |0_a\rangle - i\sin\left(c_1\delta\tau\right)\sigma_z^1|\psi_\mathcal{S}\rangle \otimes \sigma_x^a|0_a\rangle \\ &= \cos\left(c_1\delta\tau\right)|\psi_\mathcal{S}\rangle \otimes |0_a\rangle - i\sin\left(c_1\delta\tau\right)\sigma_z^1|\psi_\mathcal{S}\rangle \otimes |1_a\rangle \end{aligned} \quad (26)$$

The next step of the algorithm is the reset, *i.e.*, measure and reinitialization of the ancilla. Since the ancilla and the system are in an entangled state, the measurement of the ancilla affects the state of the system. In particular, if state 0 is measured, then the system will be in state $|\psi_\mathcal{S}\rangle$, while if state 1 is measured the system will be in $\sigma_z^1|\psi_\mathcal{S}\rangle$. These states correspond respectively to an *avoided* collision, in which the system remains in the same state, or an *occurred* collision, in which the phase of the interacting qubit (site) is flipped. In practice, the operation "interaction and reset" acts on the system-qubit as a probabilistic Z-gate which reverses the phase of the system-qubit with probability $\left|\sin\left(c_j\delta\tau\right)\right|^2$. Notice that for $\delta\tau \ll 1$, which is a condition to approximate the Lindblad master equation, the collision event is unlikely and most of the reset gates have the only effect of erasing the correlation created between the system and the ancilla qubit. Because the result of the measurement on the ancilla is stochastic at each repetition of the circuit, we devise a modified algorithm that replicates a single quantum trajectory. To do this, we simply save on a separate classical register the sequence of the outcomes obtained from the measurement of the ancilla. This is a string, let us say **b**, of $NSm$ bits, where $N$ is the number of sites, $S$ is the number of time steps and $m$ is the number of Trotter steps. Once the history of the outcomes is accessible, the same trajectory can be generated multiple times by using only the quantum register encoding the system. In the execution circuit of Figure 3b, the 2-qubit gates involving system and ancilla are then replaced with the I-gate (identity-gate), if the corresponding bit in string **b** is 0, and with the Z-gate if the bit is 1; this is shown in Figure 7.

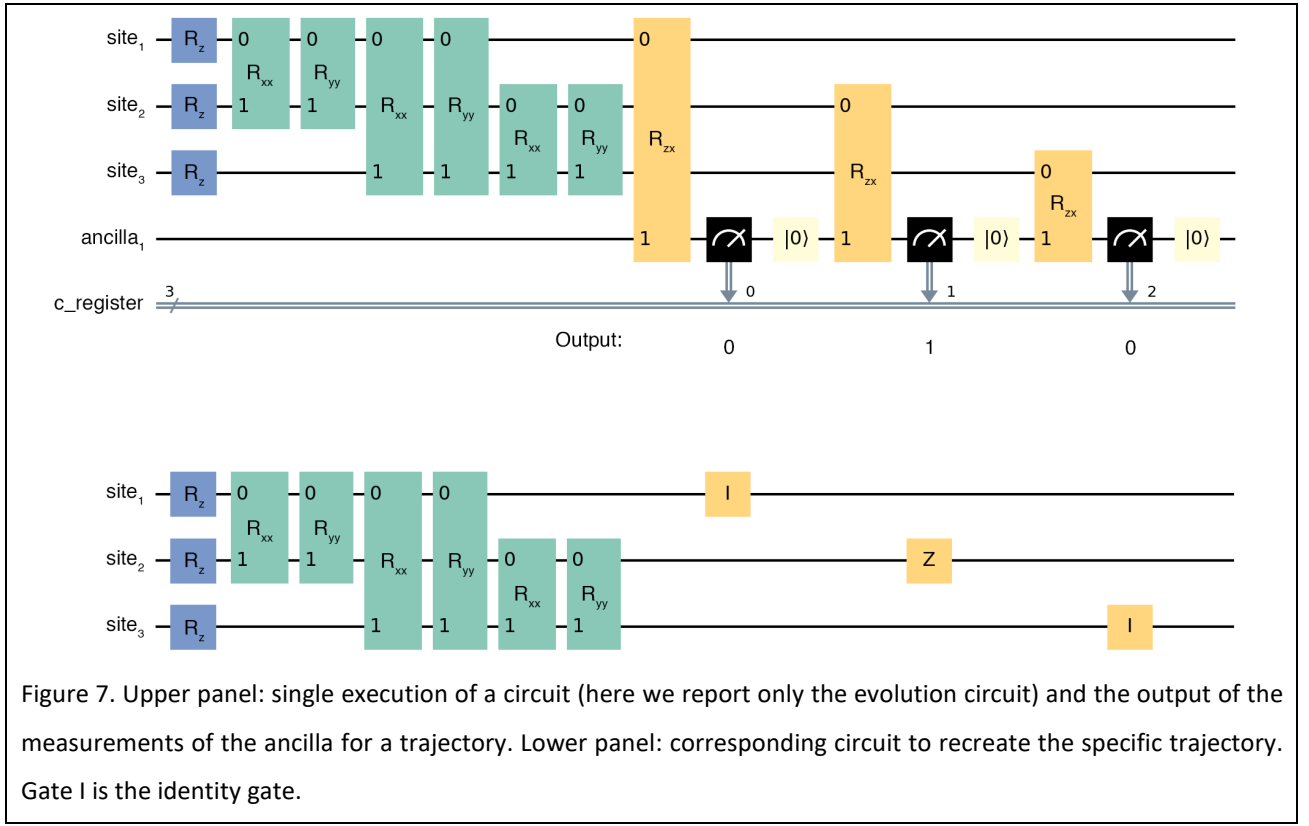

Figure 7. Upper panel: single execution of a circuit (here we report only the evolution circuit) and the output of the measurements of the ancilla for a trajectory. Lower panel: corresponding circuit to recreate the specific trajectory. Gate I is the identity gate.

Figure 8a shows a single trajectory (blue line) generated by the procedure described above together with the dynamics of the isolated system (green line). Because the probability of the system-ancilla collision is low, the two dynamics coincide for the initial time interval until an effective collision happens. The waiting time until the first effective collision has naturally a random character as we can see by looking at the swarm of trajectories reported in Figure 8b. Initially, the dynamics of the isolated system is clearly reproduced by the overlay of multiple trajectories (blue lines in transparency) and gradually fades as the probability to have observed an effective collision increases with time. The solid blue line represents the dynamics of the open system obtained as an average of 200 trajectories while the orange dotted line is the solution of the corresponding master equation. By comparing the trajectories of the collision algorithm in Figure 8 with those of the classical noise algorithm shown in Figure 2, it is evident that they generate a different distribution of the observable despite providing the same average value. Other algorithms can be devised based on a different unravelling of the master equation, for example by realizing the trajectories obtained with the traditional *quantum jump* approach to open system dynamics as used by Govia et al. in ref. [84].

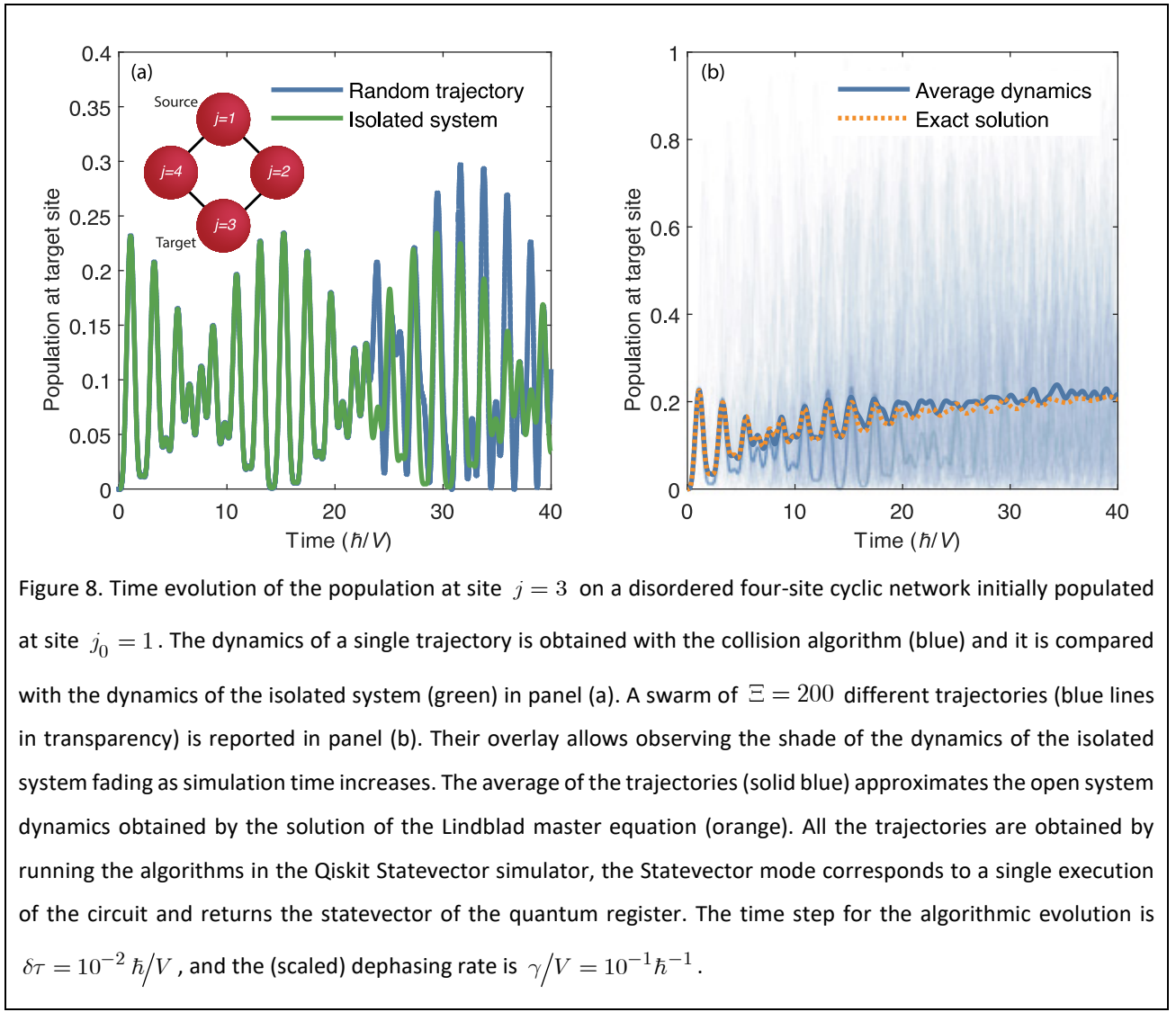

Figure 8. Time evolution of the population at site $j=3$ on a disordered four-site cyclic network initially populated at site $j_0 = 1$. The dynamics of a single trajectory is obtained with the collision algorithm (blue) and it is compared with the dynamics of the isolated system (green) in panel (a). A swarm of $\Xi = 200$ different trajectories (blue lines in transparency) is reported in panel (b). Their overlay allows observing the shade of the dynamics of the isolated system fading as simulation time increases. The average of the trajectories (solid blue) approximates the open system dynamics obtained by the solution of the Lindblad master equation (orange). All the trajectories are obtained by running the algorithms in the Qiskit Statevector simulator, the Statevector mode corresponds to a single execution of the circuit and returns the statevector of the quantum register. The time step for the algorithmic evolution is $\delta\tau = 10^{-2}\,\hbar/V$, and the (scaled) dephasing rate is $\gamma/V = 10^{-1}\hbar^{-1}$.

# 7    Conclusions

Simulating quantum dynamics in molecular networks necessarily requires taking into account the role of the environment in an effective way as it induces qualitatively different phenomena as the environment-assisted quantum transport. We have analysed in detail two different quantum algorithms for the simulation of dephasing-assisted quantum transport in digital quantum computers, focusing on the solution of the Lindblad master equation where site decoherence is included in the dissipator. The first algorithm is based on an explicit average over dynamics obtained by solving the Schrödinger equation with a stochastic Hamiltonian while the second algorithm is based on a collision scheme. Both algorithms can be applied with a physical mapping of the network, where each site is encoded in a qubit, and with an algorithmic mapping in which each site is encoded in an element of the computational basis. The circuit complexity is the same in the two cases and for both algorithms, and it scales quadratically with the number of quantum sites in the network. The two encodings provide different advantages: the algorithmic mapping allows an exponential saving of memory resources as it requires only $\lceil \log_2 N \rceil$ qubits to represent an $N$-site network. On the other hand, it entails a larger computational effort for the gate decomposition of the Hamiltonian part of the dynamics. The physical mapping does not bring a memory advantage for simulations confined to the single exciton manifold, but it can straightforwardly accommodate more excitons unleashing the quantum advantage of simulating many-body dynamics.

We test the algorithms on the Qiskit QASM simulator by calculating the dynamics and the transfer efficiency in a disordered four-site ring with different strengths of the environment-induced site decoherence. Our simulations find that an intermediate level of decoherence enhances the transport efficiency which is the hallmark of environment-assisted quantum transport. We have further analysed the two algorithms in the framework of quantum trajectories and found that quantum computers are versatile testbeds for unravelling-based implementations. In fact, while the approaches based on quantum trajectories in classical computers implies the overhead of generating a statistical sample which is not necessary in principle, quantum trajectories appear a natural tool to analyse quantum dynamical simulations because of the inherent necessity of repeating the circuit to extract expectation values.

We should point out that the specific type of open system dynamics considered in this contribution is a simple instance of the possible effect of the environment, only considering decoherence between different sites. Because of its structure, the site-dephasing master equation does not describe the effects of the temperature (indeed it corresponds to a high-temperature limit) and being in the Lindblad form, it cannot describe environments beyond the Markovian assumption underlying the semigroup theory. However, both the algorithms have been designed to include an effective representation of the environment, in the form of a stochastic process (classical noise algorithm) or additional degrees of freedom (collision algorithm). This is because we aim at algorithmic strategies which allow building different environments to simulate their effects on the dynamics of a target system. For example, the white noise scheme we have used in the classical noise algorithm can be easily generalized to implement coloured noise, for which closed equations for the average dynamics are not analytically available in general. Moreover, the collision scheme allows us to include temperature effects [85], spatial and time correlations in the environment by designing the states of the ancillae and the interaction with the system's subunits [65,86]. Therefore, this work serves as a benchmark of these algorithmic strategies by demonstrating their reliability in the simulation of a well-defined open-system dynamics whose master equation is known.

## Acknowledgements

This work was supported by the Department of Chemical Sciences (DiSC) and the University of Padova through the project QA-CHEM (P-DiSC#04BIRD2021-UNIPD). Computational work has been carried out on the C3P (Computational Chemistry Community in Padua) HPC facility of the Department of Chemical Sciences of the University of Padova. We acknowledge the use of IBM Quantum services for this work. The views expressed are those of the authors and do not reflect the official policy or position of IBM or the IBM Quantum team.

## Appendix A

We show that the classical noise algorithm approximates on average the site-dephasing master equation. The state of the system at time $s\delta\tau$ is obtained applying the evolution operator $U_{\xi,s}\left(s\delta\tau,(s-1)\delta\tau\right) = \exp\left(-iH\delta\tau - iH_{fluc,\xi,s}\sqrt{\delta\tau}\right)$ to the state at $(s-1)\delta\tau$,

$$\left|\psi(t+\delta\tau)\right\rangle\left\langle\psi(t+\delta\tau)\right| = U_{\xi,s}\left(s\delta\tau,(s-1)\delta\tau\right)\left|\psi(t)\right\rangle\left\langle\psi(t)\right|U^{\dagger}_{\xi,s}\left(s\delta\tau,(s-1)\delta\tau\right), \tag{27}$$

where $t + \delta\tau = s\delta\tau$ for simplicity. By assuming the integration step $\delta\tau$ to be small enough, we can expand the evolution operator to first order in $\delta\tau$,

$$U_{\xi,s}\left(s\delta\tau,(s-1)\delta\tau\right) \approx \mathbb{I}^{\otimes\lceil\log_2 N\rceil} - iH\delta\tau - iH_{fluc,\xi,s}\sqrt{\delta\tau} - \frac{1}{2}H^2_{fluc,\xi,s}\delta\tau. \tag{28}$$

Thus, eq. (27) can be written as

$$\begin{aligned}|\psi(t+\delta\tau)\rangle\langle\psi(t+\delta\tau)| \approx & |\psi(t)\rangle\langle\psi(t)|+ \\ & -i\big[H,|\psi(t)\rangle\langle\psi(t)|\big]\delta\tau - i\big[H_{fluc,\xi,s},|\psi(t)\rangle\langle\psi(t)|\big]\sqrt{\delta\tau} + \\ & +H_{fluc,\xi,s}|\psi(t)\rangle\langle\psi(t)|H_{fluc,\xi,s}\delta\tau - \frac{1}{2}\big[H^2_{fluc,\xi,s},|\psi(t)\rangle\langle\psi(t)|\big]_+\delta\tau\end{aligned} \quad (29)$$

where we have used $H_{fluc,\xi,s} = H^\dagger_{fluc,\xi,s}$ to simplify the notation.

Recalling that $H_{fluc,\xi,s} = \sum_{j=1}^{N}\delta\varepsilon_{j,\xi,s}|j\rangle\langle j| = \sum_{j=1}^{N}\delta\varepsilon_{j,\xi,s}L_j$ and $\rho(t) = \overline{|\psi(t)\rangle\langle\psi(t)|}$, by taking the average over the realizations of the stochastic process, one obtains

$$\begin{aligned}\rho(t+\delta\tau) \approx & \rho(t)+ \\ & -i\big[H,\rho(t)\big]\delta\tau - i\sum_{j=1}^{N}\Big[L_j,\overline{\delta\varepsilon_{j,\xi,s}|\psi(t)\rangle\langle\psi(t)|}\Big]\sqrt{\delta\tau} + \\ & +\sum_{j=1}^{N}L_j\overline{\delta\varepsilon^2_{j,\xi,s}|\psi(t)\rangle\langle\psi(t)|}L_j\delta\tau - \frac{1}{2}\sum_{j=1}^{N}\Big[L_jL_j,\overline{\delta\varepsilon^2_{j,\xi,s}|\psi(t)\rangle\langle\psi(t)|}\Big]_+\delta\tau\end{aligned} \quad (30)$$

Because of the statistical properties of the white noise, the following relations hold

$$\overline{\delta\varepsilon_{j,\xi,s}|\psi(t)\rangle\langle\psi(t)|} = \overline{\delta\varepsilon_{j,\xi,s}}\,\rho(t) = 0,$$

$$\overline{\delta\varepsilon^2_{j,\xi,s}|\psi(t)\rangle\langle\psi(t)|} = \overline{\delta\varepsilon^2_{j,\xi,s}}\,\rho(t) = \omega_j^2\rho(t).$$

Thus, eq. (30) reduces to the following difference quotient

$$\frac{\rho(t+\delta\tau) - \rho(t)}{\delta\tau} \approx -i\big[H,\rho(t)\big] + \sum_{j=1}^{N}\omega_j^2\left(L_j\rho(t)L_j - \frac{1}{2}\big[L_jL_j,\rho(t)\big]_+\right), \quad (31)$$

that gives the master equation (7) in the limit of $\delta\tau \to 0^+$.

### Appendix B

In its most common microscopic derivation, the Lindblad dynamics of an open quantum system is generated by a weak interaction with a fast thermal environment [17]. The environment is assumed to be large enough so that the thermal state is not affected by the interaction with the system and its relaxation time is assumed to be shorter than the typical evolution of the system due to the interaction with the bath. All microscopic derivations of Markovian master equations implicitly set a characteristic time under which the needed assumptions are violated. As such quantum Markovian dynamics imply a *coarse-graining* of time. To derive the master equation from a collision scheme, we identify the time scale of the coarse-graining $\delta\tau$ with the collision time. In the limit of small $\delta\tau$, we can expand the time evolution operator $U = \exp(-iH_{CM}\delta\tau)$ as

$$U \approx \mathbb{I}^{\otimes N} \otimes \mathbb{I}^{\otimes N} - iH_{ex}\otimes\mathbb{I}^{\otimes N}\delta\tau - iH_{int}\delta\tau - \frac{1}{2}H^2_{int}\delta\tau^2. \quad (32)$$

The need of expanding the interaction Hamiltonian to second-order will become clear in the following. By substituting the above expression into the dynamical map in eq. (21) and keeping linear terms in $H_{ex} \otimes \mathbb{I}^{\otimes N}$ and quadratic terms in $H_{int}$, one gets

$$\begin{aligned}\rho_S(t+\delta\tau) \approx \rho_S(t) + \\ -i\operatorname{Tr}_a\left\{\left[H_{ex}\otimes\mathbb{I}^{\otimes N},\rho_S(t)\otimes\rho_a\right]\right\}\delta\tau - i\operatorname{Tr}_a\left\{\left[H_{int},\rho_S(t)\otimes\rho_a\right]\right\}\delta\tau + \\ +\operatorname{Tr}_a\left\{H_{int}\left(\rho_S(t)\otimes\rho_a\right)H_{int}^\dagger\right\}\delta\tau^2 - \frac{1}{2}\operatorname{Tr}_a\left\{\left[H_{int}^2,\rho_S(t)\otimes\rho_a\right]_+\right\}\delta\tau^2\end{aligned} \quad (33)$$

By assuming that the density matrix of the ancilla subspace is diagonal in the computational basis, the following equalities hold

$$\operatorname{Tr}_a\left\{\sigma_x^{a_j}\rho_a\right\} = 0$$
$$\operatorname{Tr}_a\left\{\sigma_x^{a_j}\sigma_x^{a_{j'}}\rho_a\right\} = \begin{cases} 1 & j=j' \\ 0 & \text{otherwise}\end{cases} \quad (34)$$

and eq. (33) can be written as

$$\rho_S(t+\delta\tau) \approx \rho_S(t) - i\left[H_{ex},\rho_S(t)\right]\delta\tau + \sum_{j=1}^N c_j^2\delta\tau^2\left(\sigma_z^j\rho_S(t)\sigma_z^j - \rho_S(t)\right), \quad (35)$$

By transforming into a difference quotient during the collision time one obtains

$$\frac{\rho_S(t+\delta\tau)-\rho_S(t)}{\delta\tau} \approx -i\left[H_{ex},\rho_S(t)\right] + \sum_{j=1}^N c_j^2\delta\tau\left(\sigma_z^j\rho_S(t)\sigma_z^j - \rho_S(t)\right). \quad (36)$$

To recover the master equation, the next step is to apply the limit $\delta\tau \to 0^+$. However, if the coupling constants $c_j$ are assumed independent of $\delta\tau$, the summation on the right-hand side of eq. (36) vanishes in the limit. It is, therefore, appropriate to define the coupling terms as $c_j^2 = \Gamma_j/\delta\tau$, where $\Gamma_j$ are constants. In this way, eq. (11) is obtained. Note that, with this definition of $c_j$, the second-order expansion in eq. (32) corresponds to a first-order term in $\delta\tau$.

## Appendix C

In this Appendix we briefly discuss the collision algorithm for an algorithmic mapping. The algorithmic mapping should be applied only to the quantum register of the system since it is not advantageous in the single ancilla register. In this framework, the Hamiltonian of the system is the same as in eq. (3), while for the interaction of each site $j$ with the ancilla, the form $c_j|j\rangle\langle j|\otimes\sigma_z^a$ can be assumed, where the first operator acts on the site of the system and the second on the ancilla. Note that this form of the interaction differs from what we used for the algorithm in the physical mapping, eq. (20), because of the $\sigma_z$ on the ancilla qubit. This form is convenient because the operator associated with the site-ancilla interaction Hamiltonian becomes diagonal in the computational basis, and thus it can be decomposed into a smaller number of gates (see Section 5.1). However, to simulate the site-dephasing master equation with this form of interaction, the ancilla system must be prepared in the state $\rho_a = \left(|0\rangle\langle 0| + |1\rangle\langle 1|\right)/2$. Although this mixed state cannot be directly implemented on a quantum computer, the same effect is obtained by randomly initializing the ancilla to state $|0\rangle$ or $|1\rangle$ with probability $1/2$.

The algorithm can be divided into two main parts: the initialization of the system that is typical for the algorithmic mapping, and the dynamics through evolution gates. Assuming a small collision time $\delta\tau$, each evolution gate can be factorized into an evolution acting only on the system, $\exp(-iH\delta\tau)$, which is decomposed into $O\left(4^{\lceil \log_2 N \rceil}\right) \approx O(N^2)$ CNOT gates (see Section 5.1), and $N$ interactions between the sites and the ancilla, driven by the gate associated with operator $\exp\left(-ic_j |j\rangle\langle j| \otimes \sigma_z^a \delta\tau\right)$, each one followed by the reset and reinitialization of the state of the ancilla. The (diagonal) evolution operator of the interaction can be decomposed into $O(N)$ CNOT gates. The number of CNOT gates thus scales as $O(N^2)$ when all the interactions are taken into accounts. The collision algorithm in the algorithmic mapping thus scales logarithmically in the number of qubits required for the implementation of the system $\lceil \log_2 N \rceil$ with the addition of one ancilla qubit, and it scales quadratically with respect to the number of CNOT gates.